\documentstyle[11pt]{article}
\newcommand{\f}{{\rm f}}
\newcommand{\g}{{\rm g}}

\newcommand{\sd}{\tilde{\rm d}}
\newcommand{\su}{\tilde{\rm u}}
\newcommand{\sch}{\tilde{\rm c}}
\newcommand{\sst}{\tilde{\rm s}}
\newcommand{\st}{\tilde{\rm t}}
\newcommand{\sbo}{\tilde{\rm b}}
\newcommand{\se}{\tilde{\rm e}}
\newcommand{\smu}{\tilde{\mu}}
\newcommand{\stau}{\tilde{\tau}}
\newcommand{\snu}{\tilde{\nu}}
\newcommand{\sell}{\tilde{\ell}}
\newcommand{\snue}{\tilde{\nu}_{e}}
\newcommand{\snum}{\tilde{\nu}_{\mu}}
\newcommand{\snut}{\tilde{\nu}_{\tau}}
\newcommand{\glu}{\tilde{\rm g}}
\newcommand{\chio}{\tilde{\rm N}}
\newcommand{\chip}{\tilde{\rm C}^\pm}
\newcommand{\chim}{\tilde{\rm C}^\mp}
\newcommand{\grav}{\tilde{\rm G}}

\newcommand{\fbar}{\overline{\rm f}}

\newcommand{\sq}{\tilde{\rm q}}
\newcommand{\sqs}{\tilde{\rm q}^*}
\newcommand{\tp}{\tilde{\rm t}}
\newcommand{\tm}{\tilde{\rm t}^*}

\newcommand{\ttimss}{{\tt IMSS}}
\newcommand{\ttrmss}{{\tt RMSS}}
\newcommand{\drawbox}[1]{\vspace{\baselineskip}\noindent%
\fbox{\texttt{#1}}\vspace{0.5\baselineskip}}

\newenvironment{entry}%
{\begin{list}{}{\setlength{\topsep}{0mm} \setlength{\itemsep}{0mm}
\setlength{\parskip}{0mm} \setlength{\parsep}{0mm}
\setlength{\leftmargin}{20mm} \setlength{\rightmargin}{0mm}
\setlength{\labelwidth}{18mm} \setlength{\labelsep}{2mm}}}%
{\end{list}}
\newenvironment{subentry}%
{\begin{list}{}{\setlength{\topsep}{0mm} \setlength{\itemsep}{0mm}
\setlength{\parskip}{0mm} \setlength{\parsep}{0mm}
\setlength{\leftmargin}{10mm} \setlength{\rightmargin}{0mm}
\setlength{\labelwidth}{18mm} \setlength{\labelsep}{2mm}}}%
{\end{list}}
\newcommand{\itemc}[1]{\item[\textbf{#1}\hfill]}
\newcommand{\iteme}[1]{\item[\texttt{#1}\hfill]}

\begin{document}
\pagenumbering{arabic}
\pagestyle{plain}
%
\begin{titlepage}
\begin{flushright}
ANL--HEP--PR--96--63 \\
\end{flushright}
\vspace{1in}
\begin{center}
\large
{\tt SPYTHIA}, A Supersymmetric Extension of {\tt PYTHIA 5.7}
\begin{center}
{\bf S. Mrenna}\footnote{mrenna@hep.anl.gov}
\end{center}
\begin{center}
{High Energy Physics Division \\
Argonne National Laboratory \\
Argonne, IL  60439}
\end{center}
\end{center}
\vspace{2in}
\raggedbottom
\setcounter{page}{1}
\relax

\begin{abstract}
{\tt SPYTHIA} is an event level Monte Carlo program which simulates
particle production and decay at lepton and hadron colliders in
the Minimal Supersymmetric Standard Model (MSSM).  It is an extension
of {\tt PYTHIA 5.7}, with all of its previous capabilities.  
This paper is meant to supplement the {\tt PYTHIA/JETSET} user
manual, providing a description of the new particle spectrum,
hard scattering processes, and decay modes.  Several examples of
using the program are provided.
\end{abstract}

\vspace{2.0cm}
\end{titlepage}

\newpage

\section{Introduction}

Recently, experiments at the Tevatron collider at Fermilab and
the Large Electron Positron (LEP) collider at CERN have greatly
extended our understanding of the Standard Model (SM).
Several LEP energy upgrades
have been staged on a migration towards LEP190, which will be
replaced by the Large Hadron Collider (LHC).
Prior to the operation of the LHC, the Tevatron collider will begin the
Main Injector era, possibly with an extended running time,
perhaps even at a higher luminosity than originally planned.  
Despite its phenomenological
successes, the SM suffers from various theoretical problems which
make it seem unlikely to be a complete theory.
The rich physics program outline above is expected to definitively probe 
the energy regime responsible for Electroweak Symmetry Breaking (EWSB).
One promising extension of the SM, which successfully addresses
several of its deficiencies, is the 
the Minimal Supersymmetric Standard Model (MSSM) \cite{baer}.  
Monte Carlo simulations are useful tools for studying 
the phenomenological implications of a new theory, particularly
for determining search strategies and optimizing detector design.
{\tt SPYTHIA} is an event level Monte Carlo program which simulates
particle production and decay at lepton and hadron colliders in
the MSSM.  It is an extension
of {\tt PYTHIA 5.7} \cite{pyt57}, with all of its previous capabilities,
and additional particles, hard scattering processes, and decays.
Furthermore, the simulation of the MSSM Higgs sector already
present in {\tt PYTHIA} is extended to include the decay of Higgs bosons 
into MSSM particles and the decay of MSSM particles into Higgs boson.

The are already a few similar programs available.
Most of the processes included in {\tt SPYTHIA}
are already present in the {\tt ISAJET/ISASUSY} \cite{isajet}
event generator.  In addition, a number of processes and decays relevant for
lepton colliders are included in the generator {\tt
SUSYGEN} \cite{susygen}, which
is interfaced to {\tt JETSET} \cite{jetset} and includes initial 
state photon
radiation.  The development of {\tt SPYTHIA} arose from the desire
to study the phenomenology of the MSSM at lepton--lepton,
lepton--hadron, and hadron--hadron colliders using the 
initial and final state radiation and fragmentation models of
{\tt PYTHIA/JETSET}.
Additionally, one can perform cross checks and estimate model
uncertainties with other generators.

Like the SM, the MSSM contains a number of parameters with the
dimension of mass which are not fixed by the theory.  Supergravity
(SUGRA) inspired models reduce the number of free parameters by
imposing universality and exploiting the apparent unification of
gauge couplings.  Five parameters fixed at the gauge coupling
unification scale, 
$\tan\beta, M_0, m_{1/2}, A_0,$ and $sign(\mu)$, are then related
to the mass parameters at the scale of EWSB by renormalization
group equations \cite{pierce}.  {\tt ISASUSY} and {\tt SUSYGEN} 
numerically solve these equations to determine the mass
parameters.  Alternatively, they can input a general set of
parameters.  {\tt SPYTHIA} operates in the second manner,
with a slightly more general set of input parameters.  There are
three reasons for this:  (1) programs already exist to calculate
the SUGRA inspired mass parameters, (2) approximate analytic
formulae \cite{drees} also exist which reproduce the output of
{\tt ISASUSY} within $\simeq 10\%$, and (3) we desire to study a
much richer phenomenology than that possible in SUGRA inspired
models.  The {\tt SPYTHIA} input parameters are described in detail
later.

In the following, it is assumed that the reader has a working knowledge of
{\tt PYTHIA/JETSET}\footnote{A postscript version of the user manual
for {\tt PYTHIA 5.7/JETSET 7.4}
is available at the {\tt URL} address
http://thep.lu.se/tf2/staff/torbjorn/manual.ps.}.
The modifications made in {\tt SPYTHIA} are the sole product 
and responsibility of this author and not H.U.~Bengtsson or
T.~Sj\"ostrand.
However, these modifications will be included in the future release
of {\tt PYTHIA 6.1}.
Sec. 2 explains the physics assumptions behind the implementation of
supersymmetry in {\tt PYTHIA} and catalogs the new particles,
processes and input parameters.  
The parameters and routines used in the MSSM simulation are described in Sec. 3.
Availability and setup of the {\tt SPYTHIA} package are
detailed in Sec.~4, along with some examples of its use.  
Conclusions are presented in Sec.~5.

\section{Simulation of Supersymmetry}

\subsection{Particle Spectrum}
{\tt SPYTHIA} assumes the particle content of the MSSM.
Each SM fermion has a scalar partner with the same quantum
numbers.  For counting purposes, the fermion states
$\psi_L$ and $\psi_R$ are separate particles
with scalar partners $\phi_L$ and $\phi_R$.  Each SM gauge boson has
a fermion partner with the same quantum numbers.  The Higgs sector is
extended to two complex scalar doublets, leading to 3 additional 
physical Higgs bosons ${\rm H,A}$ and ${\rm H}^\pm$ to complement the SM 
Higgs ${\rm h}$.
Fermion higgsino partners are added for the two scalar doublets.
Additionally, a light gravitino $\grav$ is included to allow
studies of gauge mediated dynamical SUSY breaking \cite{dine}.
The particle partners and {\tt SPYTHIA KF} codes are listed in 
Table~1.  Note that antiparticles of scalar particles are denoted 
by $^*$ in the text.

Because of the large Yukawa 
couplings and the running masses for colored particles, the interaction and 
mass eigenstates for the third generation sfermions can be 
significantly mixed.  We denote the top and bottom squark mass 
eigenstates $\st_1$,$\st_2$,$\sbo_1$, and $\sbo_2$ to distinguish them from 
the nearly degenerate eigenstates for the lighter squarks.  In addition,
the tau slepton mass eigenstates are $\tilde\tau_1$ and $\tilde\tau_2$.  
The subscript $1$ or $2$ refers to the lightest and heaviest state
respectively.  In SUGRA inspired models, in the absence of mixing
so that interaction eigenstates are the same as mass eigenstates, the
right eigenstate is lighter than the left.  In this case, for example,
$\sbo_1 = \sbo_R$ and $\sbo_2 = \sbo_L$.
For completeness, we include the sterile $\snu_R$ particles.  
The mixing of the partners to 
the electroweak gauge bosons (gauginos) and the two Higgs doublets 
(higgsinos) lead to the mass 
eigenstates for neutralino $\tilde{\rm N}_i$ and 
chargino $\tilde{\rm C}_i$ particles.  
A complete listing of the full particle spectrum, 
including decay channels, branching ratios, and the total width
is available (as usual) through the subroutine {\tt LULIST}.

\begin{table}[pt]
\caption{{\tt SPYTHIA} particle {\tt KF} codes.}

\begin{center}
\begin{tabular}{|c|c|c||c|c|c|}
\hline
{\tt KF} & Name & Printed & {\tt KF} & Name & Printed  \\
\hline
   41    & $\sd_L$ & \~{}{\tt d\_L}       & 59 & $\snum{}_L$ & \~{}{\tt numu\_L} \\    
   42    & $\sd_R$ & \~{}{\tt d\_R}       & 60 & $\snum{}_R$ & \~{}{\tt numu\_R} \\    
   43    & $\su_L$ & \~{}{\tt u\_L}       & 61 & $\stau{}_1$ & \~{}{\tt tau\_1} \\     
   44    & $\su_R$ & \~{}{\tt u\_R}       & 62 & $\stau{}_2$ & \~{}{\tt tau\_2} \\     
   45    & $\sst_1$ & \~{}{\tt s\_L}      & 63 & $\snut{}_L$ & \~{}{\tt nutau\_L} \\   
   46    & $\sst_2$ & \~{}{\tt s\_R}      & 64 & $\snut{}_R$ & \~{}{\tt nutau\_R} \\   
   47    & $\sch_L$ & \~{}{\tt c\_L}      & 65 & $\glu$    & \~{}{\tt g} \\     
   48    & $\sch_R$ & \~{}{\tt c\_R}      & 66 & $\chio_1$ & \~{}{\tt {N}$_1$} \\          
   49    & $\sbo_1$ & \~{}{\tt b\_1}      & 67 & $\chio_2$ & \~{}{\tt {N}$_2$} \\          
   50    & $\sbo_2$ & \~{}{\tt b\_2}      & 68 & $\chio_3$ & \~{}{\tt {N}$_3$} \\          
   51    & $\st_1$ & \~{}{\tt t\_1}       & 69 & $\chio_4$ & \~{}{\tt {N}$_4$} \\          
   52    & $\st_2$ & \~{}{\tt t\_2}       & 70 & $\chip_1$ & \~{}{\tt C$_1$} \\          
   53    & $\se_L$ & \~{}{\tt e\_L}       & 71 & $\chip_2$ & \~{}{\tt C$_2$} \\          
   54    & $\se_R$ & \~{}{\tt e\_R}       &   &  & \\                
   55    & $\snue{}_L$ & \~{}{\tt nue\_L}   &   & & \\                
   56    & $\snue{}_R$ & \~{}{\tt nue\_R}   &   & & \\                
   57    & $\smu{}_L$ & \~{}{\tt mu\_L}     &   & & \\           
   58    & $\smu{}_R$ & \~{}{\tt mu\_R}     & 76 & $\grav$ & \~{}{\tt grvtno} \\   
\hline
\end{tabular}
\end{center}
\end{table}

\subsection{Particle Decays}
In {\tt PYTHIA}, resonances are decayed so 
that color flow information is correctly passed from unstable particles
to their decay products and no unstable colored 
particles are passed on to {\tt JETSET}.  
In {\tt SPYTHIA}, all Supersymmetric partners are
treated as resonances.
{\tt SPYTHIA} assumes R--symmetry, which has two major phenomenological 
consequences:  
(1) Supersymmetric particles are produced in pairs,
and (2) there is one and only one stable Supersymmetric particle.  
By default, the lightest
superpartner (LSP) is $\chio_1$, unless $\chio_1$ is allowed to decay to
a gravitino.  In that case, $\chio_1$ is the next to lightest
superpartner (NLSP) and $\grav$ is the LSP.

The decays of superpartners are calculated using the formulae of
Refs.~\cite{gunion1,bartl1,bartl2,bartl3}.  
All decays are spin averaged.
For simplicity, the kinematics of 
three body decays of neutralinos, charginos, and the gluino are sampled 
using only the phase space weight.  Decays involving $\sbo$ and $\st$ use the
formulae of \cite{bartl3}, so they are valid for large values of
$\tan\beta$.  The one loop decays $\chio_j\to\chio_i\gamma$ and 
$\st\to c\chio_1$ are also included.

\subsection{Processes}
\indent Tables~2--5 are meant to update 
Tables~11--14 in the {\tt Pythia 5.7}
manual.  The number of available processes in {\tt SPYTHIA} 
is extended to 400, with processes 201--280 reserved for
the MSSM.  In processes 210 and 213, $\sell$ refers to both $\se$ and
$\smu$.  For ease of readability, we have removed the subscript $L$ on $\snu$.
$\tp_i\tm_i, \stau_i\stau_j^*$ and $\stau_i\snu_\tau^*$ 
production correctly account for sfermion mixing.  Several processes
are conspicuously absent from the table.  For example, processes
{\tt 255} and {\tt 257} would simulate the associated production 
of right handed squarks with charginos.  Since the right handed squark
only couples to the higgsino component of the chargino, the interaction
strength is proportional to the quark mass, so these processes can
be ignored.   

\begin{table}[hpt]
\caption{Subprocess codes, part 1. First column is `+' for processes
implemented and blank for those that are only foreseen. Second is
the subprocess number {\tt ISUB}, and third the description of the
process. The final column gives references from which the
cross sections have been obtained.}

\begin{center}
\begin{tabular}{|c|r|l|l|}
\hline
In & No. & Subprocess & Reference \\
\hline
+  & 201 & $\f_i \fbar_i \to \se_L \se_L^*$ & \cite{bartl4,dawson1} \\
+  & 202 & $\f_i \fbar_i \to \se_R \se_R^*$ & \cite{bartl4,dawson1} \\
+   & 203 & $\f_i \fbar_i \to \se_L \se_R^*+\se_L^* \se_R$ &
\cite{bartl4} \\
+  & 204 & $\f_i \fbar_i \to \smu_L \smu_L^*$ & \cite{bartl4,dawson1} \\
+  & 205 & $\f_i \fbar_i \to \smu_R \smu_R^*$ & \cite{bartl4,dawson1} \\
+   & 206 & $\f_i \fbar_i\to\smu_L \smu_R^*+\smu_L^* \smu_R$ &
\cite{bartl4} \\
+  & 207 & $\f_i \fbar_i\to\stau_1 \stau_1^*$ & \cite{bartl4,dawson1} \\
+  & 208 & $\f_i \fbar_i\to\stau_2 \stau_2^*$ & \cite{bartl4,dawson1} \\
+   & 209 & $\f_i \fbar_i\to\stau_1
\stau_2^*+\stau_1^*\stau_2$&\cite{bartl4} \\
+  & 210 & $\f_i \fbar_j\to \sell_L {\snu}_\ell^*+
\sell_L^* \snu_\ell$&\cite{dawson1} \\
+  & 211 & $\f_i \fbar_j\to \stau_1
\tilde{\nu}_\tau^*+\stau_1^*\tilde{\nu}_\tau$ & \cite{dawson1} \\
+  & 212 & $\f_i \fbar_j\to \stau_2
\tilde{\nu}_\tau{}^*+\stau_2^*\tilde{\nu}_\tau$ 
& \cite{dawson1} \\
+  & 213 & $\f_i \fbar_i\to \tilde{\nu_\ell} \tilde{\nu_\ell}^*$ & 
\cite{bartl4,dawson1} \\
+  & 214 & $\f_i \fbar_i\to \tilde{\nu}_{\tau} \tilde{\nu}_{\tau}^*$ 
& \cite{bartl4,dawson1} \\
+ & 216 & $\f_i \fbar_i \to \chio_1 \chio_1$ & \cite{bartl1} \\
+ & 217 & $\f_i \fbar_i \to \chio_2 \chio_2$ & \cite{bartl1} \\
+ & 218 & $\f_i \fbar_i \to \chio_3 \chio_3$ & \cite{bartl1} \\
+ & 219 & $\f_i \fbar_i \to \chio_4 \chio_4$ & \cite{bartl1} \\
+ & 220 & $\f_i \fbar_i \to \chio_1 \chio_2$ & \cite{bartl1} \\
+ & 221 & $\f_i \fbar_i \to \chio_1 \chio_3$ & \cite{bartl1} \\
+ & 222 & $\f_i \fbar_i \to \chio_1 \chio_4$ & \cite{bartl1} \\
+ & 223 & $\f_i \fbar_i \to \chio_2 \chio_3$ & \cite{bartl1} \\
+ & 224 & $\f_i \fbar_i \to \chio_2 \chio_4$ & \cite{bartl1} \\
+ & 225 & $\f_i \fbar_i \to \chio_3 \chio_4$ & \cite{bartl1} \\
+ & 226 & $\f_i \fbar_i \to \chip_1 \chim_1$ & \cite{bartl2} \\
+ & 227 & $\f_i \fbar_i \to \chip_2 \chim_2$ & \cite{bartl2} \\
+ & 228 & $\f_i \fbar_i \to \chip_1 \chim_2$ & \cite{bartl2} \\
+ & 229 & $\f_i \fbar_j \to \chio_1 \chip_1$ & \cite{bartl1,bartl2} \\
+ & 230 & $\f_i \fbar_j \to \chio_2 \chip_1$ & \cite{bartl1,bartl2} \\
+ & 231 & $\f_i \fbar_j \to \chio_3 \chip_1$ & \cite{bartl1,bartl2} \\
+ & 232 & $\f_i \fbar_j \to \chio_4 \chip_1$ & \cite{bartl1,bartl2} \\
+ & 233 & $\f_i \fbar_j \to \chio_1 \chip_2$ & \cite{bartl1,bartl2} \\
+ & 234 & $\f_i \fbar_j \to \chio_2 \chip_2$ & \cite{bartl1,bartl2} \\
+ & 235 & $\f_i \fbar_j \to \chio_3 \chip_2$ & \cite{bartl1,bartl2} \\
+ & 236 & $\f_i \fbar_j \to \chio_4 \chip_2$ & \cite{bartl1,bartl2} \\
\hline
\end{tabular}
\end{center}
\end{table}

\begin{table}[hpt]
\caption{Subprocess codes, part 2. First column is `+' for processes
implemented and blank for those that are only foreseen. Second is
the subprocess number {\tt ISUB}, and third the description of the
process. The final column gives references from which the
cross sections have been obtained.}

\begin{center}
\begin{tabular}{|c|r|l|l|}
\hline
In & No. & Subprocess & Reference \\
\hline
+ & 237 & $\f_i \fbar_i \to \glu \chio_1$ & \cite{dawson1}     \\
+ & 238 & $\f_i \fbar_i \to \glu \chio_2$ & \cite{dawson1}     \\
+ & 239 & $\f_i \fbar_i \to \glu \chio_3$ & \cite{dawson1}     \\
+ & 240 & $\f_i \fbar_i \to \glu \chio_4$ & \cite{dawson1}     \\
+ & 241 & $\f_i \fbar_j \to \glu \chip_1$ & \cite{dawson1}     \\
+ & 242 & $\f_i \fbar_j \to \glu \chip_2$ & \cite{dawson1}     \\
+ & 243 & $\f_i \fbar_i \to \glu \glu$ & \cite{dawson1} \\
+ & 244 & $\g \g \to \glu \glu$ & \cite{dawson1} \\
+ & 246 & $\f_i \g \to {\sq_i}{}_L \chio_1$ & \cite{dawson1} \\
+ & 247 & $\f_i \g \to {\sq_i}{}_R \chio_1$ & \cite{dawson1} \\
+ & 248 & $\f_i \g \to {\sq_i}{}_L \chio_2$ & \cite{dawson1} \\
+ & 249 & $\f_i \g \to {\sq_i}{}_R \chio_2$ & \cite{dawson1} \\
+ & 250 & $\f_i \g \to {\sq_i}{}_L \chio_3$ & \cite{dawson1} \\
+ & 251 & $\f_i \g \to {\sq_i}{}_R \chio_3$ & \cite{dawson1} \\
+ & 252 & $\f_i \g \to {\sq_i}{}_L \chio_4$ & \cite{dawson1} \\
+ & 253 & $\f_i \g \to {\sq_i}{}_R \chio_4$ & \cite{dawson1} \\
+ & 254 & $\f_i \g \to {\sq_j}{}_L \chip_1$ & \cite{dawson1} \\
+ & 256 & $\f_i \g \to {\sq_j}{}_L \chip_2$ & \cite{dawson1} \\
+ & 258 & $\f_i \g \to {\sq_i}{}_L \glu$ & \cite{dawson1}\\
+ & 259 & $\f_i \g \to {\sq_i}{}_R \glu$ & \cite{dawson1}\\
+ & 261 & $\f_i \fbar_i \to \tp_1 \tm_1$ & \cite{dawson1} \\
+ & 262 & $\f_i \fbar_i \to \tp_2 \tm_2$ & \cite{dawson1} \\
+ & 263 & $\f_i \fbar_i \to \tp_1 \tm_2+\tm_1 \tp_2$ & \cite{dawson1} \\
+ & 264 & $\g \g \to \tp_1 \tm_1$ & \cite{dawson1} \\
+ & 265 & $\g \g \to \tp_2 \tm_2$ & \cite{dawson1} \\
+ & 271 & $\f_i \f_j \to {\sq_i}{}_L {\sq_j}{}_L$ & \cite{dawson1} \\
+ & 272 & $\f_i \f_j \to {\sq_i}{}_R {\sq_j}{}_R$ & \cite{dawson1} \\
+ & 273 & $\f_i \f_j \to {\sq_i}{}_L {\sq_j}{}_R+
{\sq_i}{}_R {\sq_j}{}_L$ & \cite{dawson1} \\
+ & 274 & $\f_i \fbar_j \to {\sq_i}{}_L {\sqs_j}{}_L$ & \cite{dawson1} \\
+ & 275 & $\f_i \fbar_j \to {\sq_i}{}_R {\sqs_j}{}_R$ & \cite{dawson1} \\
+ & 276 & $\f_i \fbar_j \to {\sq_i}{}_L {\sqs_j}{}_R+
{\sq_i}{}_R {\sqs_j}{}_L$ & \cite{dawson1} \\
+ & 277 & $\f_i \fbar_i \to {\sq_j}{}_L {\sqs_j}{}_L$ & \cite{dawson1} \\
+ & 278 & $\f_i \fbar_i \to {\sq_j}{}_R {\sqs_j}{}_R$ & \cite{dawson1} \\
+ & 279 & $\g \g \to {\sq_i}{}_L {\sqs_i}{}_L$ & \cite{dawson1} \\
+ & 280 & $\g \g \to {\sq_i}{}_R {\sqs_i}{}_R$ & \cite{dawson1} \\
\hline
\end{tabular}
\end{center}
\end{table}

Because there are so many processes involved, there are shortcuts
to allow the simulation of various classes of signals.  These classes are
accessible through the parameter {\tt MSEL}, and are listed in Table~6.

\begin{table}[hpt]
\caption{Classes of processes accessible through the parameter {\tt MSEL}
and the individual processes codes {\tt ISUB}.}
\begin{center}
\begin{tabular}{|c|l|}
\hline
{\tt MSEL} & Description \\
\hline
39 & All MSSM processes except Higgs production \\
40 & Squark and gluino production, {\tt ISUB} = 243, 244, 258, 259, 271--280 \\
41 & Stop pair production, {\tt ISUB} = 261--265 \\
42 & Slepton pair production, {\tt ISUB} = 201--214 \\
43 & Squark or gluino with chargino or neutralino, {\tt ISUB} = 237--242, 246--256 \\
44 & Chargino--neutralino pair production, {\tt ISUB} = 216--236 \\
\hline
\end{tabular}
\end{center}
\end{table}
\vskip 10cm
\section{The Parameters and Routines of the MSSM Simulation}
All of the Supersymmetric extensions of the code are included
in the standard {\tt PYTHIA} library.  
By default, Supersymmetry is not simulated.
However, by setting various parameters, a rich MSSM phenomenology is
available.

The parameters available to the user are stored in the
{\tt FORTRAN} common block \\
{\tt COMMON/PYMSSM/IMSS(0:99),RMSS(0:99)}.
In general, options are set by the {\tt IMSS} array, while 
real valued parameters are set by {\tt RMSS}.  The entries
{\tt IMSS(0)} and {\tt RMSS(0)} are not used, but are available
for compatibility with the {\tt C} programming language.
The arrays are described below.  The default values are denoted
by (D). \\

\subsection{The MSSM Parameters}

\drawbox{COMMON/PYMSSM/IMSS(0:99),RMSS(0:99)}%
\begin{entry}
\itemc{Purpose:} to give access to parameters that allow the
simulation of the MSSM.

\iteme{IMSS(1) :} (D=0) level of MSSM simulation.
\begin{subentry}
\iteme{= 0 :} No MSSM simulation.
\iteme{= 1 :} A general MSSM simulation.  The parameters of the model
are set by the array \ttrmss.
\iteme{= 2 :} An approximate SUGRA simulation using the analytic
formulate of \cite{drees} to reduce the number of free parameters.
In this case, only five input parameters are used.
\ttrmss{\tt(1)} is the common gaugino mass
$m_{1/2}$, \ttrmss{\tt(8)} is the common scalar mass $m_0$, 
\ttrmss{\tt(4)} fixes the sign of the higgsino mass $\mu$, 
\ttrmss{\tt(16)} is the common trilinear coupling $A$, and
\ttrmss{\tt(5)} is $\tan\beta=v_2/v_1$.
\end{subentry}

\iteme{IMSS(2) :} (D=0) treatment of $U(1), SU(2),$ and $SU(3)$ gaugino
mass parameters.
\begin{subentry}
\iteme{= 0 :} The gaugino parameters $M_1, M_2$ and $M_3$ are 
 set by {\tt RMSS(1), RMSS(2),} and {\tt RMSS(3)}, i.e. there is 
 no forced relation between them.
\iteme{= 1 :} The gaugino parameters are fixed by the relation 
$M_1/\alpha_1 3/5=M_2/\alpha_2=M_3/\alpha_3=X$ and the 
parameter {\tt RMSS(1)}.  If {\tt IMSS(1)=2}, then
{\tt RMSS(1)} is treated as the common gaugino mass $m_{1/2}$ 
and {\tt RMSS(20)} is the GUT scale coupling constant
$\alpha_{GUT}$, so that $X=m_{1/2}/\alpha_{GUT}$.
\iteme{= 2 :} $M_1$ is set by {\tt RMSS(1)}, $M_2$ by {\tt RMSS(2)} 
and $M_3 = M_2\alpha_3/\alpha_2$. In such a scenario, 
the U(1) gaugino mass behaves anomalously. 
\end{subentry}

\iteme{IMSS(3) :} (D=0) treatment of the gluino mass parameter.
\begin{subentry}
\iteme{= 0 :}  The gluino mass parameter $M_3$ is used to calculate 
the gluino pole mass with the formulae of \cite{martin1}.  The effects of 
squark loops can significantly shift the mass.  
\iteme{= 1 :}  $M_3$ is the gluino pole mass.   The effects of squark
loops are assumed to have been included in this value. 
\end{subentry}

\iteme{IMSS(4) :} (D=1) treatment of the Higgs sector.
\begin{subentry}
\iteme{= 0 :}  The Higgs sector is determined by 
the approximate formulae of \cite{carena}  and the pseudoscalar mass $M_A$ 
set by {\tt RMSS(19)}.
\iteme{= 1 :} The Higgs sector is determined by the exact
formulae of \cite{carena} and the pseudoscalar mass $M_{\rm A}$ set by 
{\tt RMSS(19)}.  The pole mass for $M_{\rm A}$ is not the same as the input
parameter . 
\iteme{= 2 :} The Higgs sector is fixed by the mixing angle $\alpha$ 
set by {\tt RMSS(18)} and the mass values {\tt PMAS(I,1)}, where 
{\tt I=25,35,36,} and {\tt 37}. 
\end{subentry}

\iteme{IMSS(7) :} (D=0) treatment of the scalar masses in an
extension of SUGRA models.  The presence of additional $U(1)$
symmetries at high energy scales can modify the boundary
conditions for the scalar masses at the unification scale.
\begin{subentry}
\iteme{= 0 :}   No additional $D$--terms are included.  In
SUGRA models, all scalars have the mass $m_0$ at the 
unification scale.
\iteme{= 1 :}   {\tt RMSS}{\tt(23--25)} are the values of $D_X, D_Y$ and
$D_S$ at the unification scale in the model of \cite{martin2}.
The boundary conditions for the scalar masses are shifted based
on their quantum numbers under the additional $U(1)$ symmetries. 
\end{subentry}

\iteme{IMSS(8) :} (D=1) treatment of the $\stau$ mass eigenstates.
\begin{subentry}
\iteme{= 0 :}   The $\stau$ mass eigenstates are calculated 
using the parameters {\tt RMSS(13,14,17)}. 
\iteme{= 1 :}   The $\stau$ mass eigenstates are identical to 
the interaction eigenstates, so they are treated
identically to $\se$ and $\smu$ .
\end{subentry}

\iteme{IMSS(9) :} (D=0) treatment of the right handed squark mass
eigenstates for the first two generations.
\begin{subentry}
\iteme{= 0 :}   The $\sq_R$ masses are fixed by \ttrmss{\tt(9)}.  $\sd_R$ and
$\su_R$ are identical except for Electroweak $D$--term contributions. 
\iteme{= 1 :}   The masses of $\sd_R$ and $\su_R$ 
are fixed by {\tt RMSS(9)} and {\tt RMSS(22)} respectively. 
\end{subentry}

\iteme{IMSS(10) :} (D=0) allowed decays for $\chio_2$.
\begin{subentry}
\iteme{= 0 :}   The second lightest neutralino $\chio_2$ decays 
with a branching ratio calculated from the MSSM parameters.  
\iteme{= 1 :}  $\chio_2$ is forced to decay only to $\chio_1 \gamma$, 
regardless of the actual branching ratio. 
This can be used for detailed studies of this particular final state. 
\end{subentry}

\iteme{IMSS(11) :} (D=0) choice of the lightest superpartner (LSP).
\begin{subentry}
\iteme{= 0 :} $\chio_1$ is the LSP.  
\iteme{= 1 :} $\chio_1$ is the next to lightest superparter (NLSP) 
and the gravitino is the LSP.  The gravitino 
decay length is calculated from the gravitino 
mass set by {\tt RMSS(21)} and the $\chio_1$ mass and mixings. 
\end{subentry}

\iteme{RMSS(1) :} If \ttimss{\tt(1)=1} $M_1$, the U(1) gaugino mass.  
If \ttimss{\tt(1)=2}, then the common gaugino mass $m_{1/2}$.

\iteme{RMSS(2) :}     $M_2$, the SU(2) gaugino mass.  

\iteme{RMSS(3) :}     $M_3$, the SU(3) (gluino) mass parameter.  

\iteme{RMSS(4) :}     $\mu$, the higgsino mass parameter.  
If {\tt IMSS(1)=2}, only the sign of $\mu$ is used.

\iteme{RMSS(5) :}     $\tan\beta$, the ratio of Higgs expectation
values.

\iteme{RMSS(6) :}     Left slepton mass $M_{\tilde \ell_L}$.  
The sneutrino mass is fixed by a sum rule.

\iteme{RMSS(7) :}     Right slepton mass $M_{\tilde \ell_R}$. 

\iteme{RMSS(8) :}     Left squark mass $M_{\sq_L}$. If {\tt IMSS(1)=2},
the common scalar mass $m_0$. 

\iteme{RMSS(9) :}     Right squark mass $M_{\sq_R}$. $M_{\sq_R}$ 
when {\tt IMSS(9)=1}. 

\iteme{RMSS(10) :}    Left squark mass for the third
 generation $M_{\sq_L}$. 

\iteme{RMSS(11) :}    Right sbottom mass $M_{\sbo_R}$. 

\iteme{RMSS(12) :}    Right stop mass $M_{\st_R}$  If
negative, then it is assumed that $M_{\st_R}^2 < 0$.  

\iteme{RMSS(13) :}    Left stau mass $M_{\tilde \tau_L}$.  

\iteme{RMSS(14) :}    Right stau mass $M_{\tilde \tau_R}$.  
\iteme{RMSS(15) :}     Bottom trilinear coupling $A_{\rm b}$.   
\iteme{RMSS(16) :}    Top trilinear coupling $A_{\rm t}$.  If {\tt IMSS(1)=2}, 
the common trilinear coupling $A$. 
\iteme{RMSS(17) :}    Tau trilinear coupling $A_\tau$.   
\iteme{RMSS(18) :}    Higgs mixing angle $\alpha$. This is only used when  
all of the Higgs parameters are set by the user, i.e  {\tt IMSS(4)=2}.  
\iteme{RMSS(19) :}    Pseudoscalar Higgs mass parameter $M_{\rm A}$.  
\iteme{RMSS(20) :} (D=.041)   GUT scale coupling constant $\alpha_{GUT}$.
\iteme{RMSS(21) :} (D=1.0)   The gravitino mass (eV).  
\iteme{RMSS(22) :}    $\su_R$ mass when {\tt IMSS(9)=1}. 
\iteme{RMSS(23) :}    $D_X$ contribution to scalar masses when 
{\tt IMSS(7)=1} (GeV$^2$). 
\iteme{RMSS(24) :}    $D_Y$ contribution to scalar masses when 
{\tt IMSS(7)=1} (GeV$^2$). 
\iteme{RMSS(25) :}    $D_S$ contribution to scalar masses when 
{\tt IMSS(7)=1} (GeV$^2$). 
\end{entry}

\drawbox{COMMON/PYSSMT/ZMIX(4,4),UMIX(2,2),VMIX(2,2),SMZ(4),SMW(2),SFMIX(16,4)}%
\begin{entry}
\itemc{Purpose:} to provide information on the neutralino, chargino,
and sfermion mixing parameters.  The variables should not be changed
by the user.

\iteme{ZMIX(4,4) :} the neutralino mixing matrix in the Bino--neutral
Wino--Up higgsino--Down higgsino basis.

\iteme{UMIX(2,2) :} the chargino mixing matrix in the charged Wino--charged
higgsino basis.

\iteme{VMIX(2,2) :} the charged conjugate chargino 
mixing matrix in the wino--charged higgsino basis.

\iteme{SMZ(4) :} the signed masses of the neutralinos.

\iteme{SMW(2) :} the signed masses of the charginos.

\iteme{SFMIX(16,4) :} the sfermion mixing matrices $\bf T$ in the L--R basis,
identified by
the corresponding fermion, i.e. {\tt SFMIX(6,I)} is the stop mixing matrix.
The four entries for each sfermion are ${\rm T}_{11}, {\rm T}_{12}, 
{\rm T}_{21},$ and 
${\rm T}_{22}$.
\end{entry}

\subsection{The MSSM Physics Routines}
The following subroutines and functions need not be accessed by the
user, but are described for completeness.
\begin{entry}
\iteme{SUBROUTINE PYAPPS :} uses approximate analytic formulae to 
determine the full set of MSSM parameters from SUGRA inputs.

\iteme{SUBROUTINE PYGLUI :} calculates gluino decay modes.

\iteme{SUBROUTINE PYGQQB :} calculates three body decays of gluinos
into neutralinos or charginos and third generation fermions.  These
routines are valid for large values of $\tan\beta$.

\iteme{SUBROUTINE PYCJDC :} calculates the chargino decay modes.

\iteme{SUBROUTINE PYHEXT :} calculates the non--Standard Model decay
modes of the Higgs bosons.

\iteme{SUBROUTINE PYHGGM :} determines the Higgs boson mass spectrum
using several inputs.

\iteme{SUBROUTINE PYINOM :} finds the mass eigenstates and mixing
matrices for the charginos and neutralinos.

\iteme{SUBROUTINE PYMSIN :} initializes the MSSM simulation.

\iteme{SUBROUTINE PYNJDC :} calculates neutralino decay modes.

\iteme{SUBROUTINE PYPOLE :} computes the Higgs boson masses using
a renormalization group improved leading--log approximation and
two loop leading--log corrections.

\iteme{SUBROUTINE PYRNMT :} determines the running mass of the top
quark.

\iteme{SUBROUTINE PYSFDC :} calculates sfermion decay modes.

\iteme{SUBROUTINE PYSUBH :} computes the Higgs boson masses using
only renormalization group improved formulae.

\iteme{SUBROUTINE PYTBDY :} samples the phase space for three body
decays of neutralinos, charginos, and the gluino.

\iteme{SUBROUTINE PYTHRG :} computes the masses and mixing matrices of
the third generation sfermions.

\end{entry}

\section{Setup and Use of the {\tt SPYTHIA} Generator}

The {\tt SPTYHIA} code is available as a uuencoded file on the
World Wide
Web\footnote{http://www.hep.anl.gov/theory/mrenna/spythia.hmtl}.
It operates in the same fashion as the standard {\tt PYTHIA}
distribution.  See the {\tt PYTHIA} manual for a detailed 
example of its usage.
Below, we present several examples of how to simulate MSSM
phenomenology with {\tt SPYTHIA}.

\subsection{Example 1:  Light Stop}
The first example is an MSSM model with a light neutralino $\chio_1$ and a
light stop $\st_1$, so that ${\rm t}\to \st_1\chio_1$ can occur.  
The input parameters are 
\ttimss{\tt(1)=1}, \ttrmss{\tt(1)=70.}, 
\ttrmss{\tt(2)=70.}, \ttrmss{\tt(3)=225.},
\ttrmss{\tt(4)=-40.}, \ttrmss{\tt(5)=1.5}, \ttrmss{\tt(6)=100.}, 
\ttrmss{\tt(7)=125.}, \ttrmss{\tt(8)=250.}, \\
\ttrmss{\tt(9)=250.}, \ttrmss{\tt(10)=1500.}, 
\ttrmss{\tt(11)=1500.}, \ttrmss{\tt(12)=-128.}, 
\ttrmss{\tt(13)=100.}, \\
\ttrmss{\tt(14)=125.}, 
\ttrmss{\tt(15)=800.}, \ttrmss{\tt(16)=800.}, 
\ttrmss{\tt(17)=0.}, and \ttrmss{\tt(19)=400.0.}
The top mass is fixed at 175 GeV, {\tt PMAS(6,1)=175.0}.
The resulting model has $M_{\st_1}=55$ GeV and $M_{\chio_1}=38$ GeV.
{\tt IMSS(1)=1} turns on the MSSM simulation.
By default, there are no intrinsic relations between the gaugino masses,
so $M_1=70$ GeV, $M_2=70$ GeV, and $M_3=225$ GeV.  The pole mass of
the gluino is slightly higher than the parameter $M_3$, and the
decay $\tilde {\rm g}\to\st_1^*{\rm t}+\st_1\bar{\rm t}$ 
occurs almost 100\% of the time.

\subsection{Example 2:  Approximate SUGRA}
The second example is an approximate SUGRA model.  The input
parameters are \ttimss{\tt(1)=2}, \ttrmss{\tt(1)=200.},
\ttrmss{\tt(4)=1.,} 
\ttrmss{\tt(5)=10.},
\ttrmss{\tt(8)=800.}, and \ttrmss{\tt(16)=0.0}.
The resulting model has
$M_{\sd_L}=901$ GeV, $M_{\su_R}=890$ GeV, $M_{\st_1}=538$ GeV,
$M_{\se_L}=814$ GeV, $M_{\glu}=560$ GeV, $M_{\chio_1}=80$ GeV,
$M_{\chip_1}=151$ GeV, $M_{\rm h}=110$ GeV, and $M_{\rm A}=883$ GeV.  It 
corresponds to the choice $M_0$=800 GeV, $M_{1/2}=$200 GeV,
$\tan\beta=10$, $A_0=0$, and $sign(\mu)>0$.  The output is similar
to an {\tt ISASUSY} run, but there is not exact agreement.

\subsection{Example 3:  {\tt ISASUSY} Model}
The final example demonstrates how to convert the output of
an {\tt ISASUSY}\footnote{Information on the ISAJET and ISASUSY 
programs can be found on the World Wide Web at
http://wwwcn1.cern.ch/asd/cernlib/mc/isajet.html.} 
run using the same SUGRA inputs into the {\tt SPYTHIA} format.
This requires a general model with the parameters
\ttimss{\tt(1)=1}, \ttimss{\tt(3)=1}, \ttrmss{\tt(1)=83.81}, 
\ttrmss{\tt(2)=168.90}, \ttrmss{\tt(3)=581.83}, \ttrmss{\tt(4)=283.37},
\ttrmss{\tt(5)=10.}, \ttrmss{\tt(6)=813.63}, \ttrmss{\tt(7)=804.87}, 
\ttrmss{\tt(8)=917.73}, \\ {\ttrmss{\tt(9)=909.89}, 
\ttrmss{\tt(10)=772.87},
\ttrmss{\tt(11)=901.52}, \ttrmss{\tt(12)=588.33}, 
\ttrmss{\tt(13)=813.63}, \\
\ttrmss{\tt(14)=804.87}, \ttrmss{\tt(15)=610.54},
\ttrmss{\tt(16)=422.35},
\ttrmss{\tt(17)=600.}, and \\ \ttrmss{\tt(19)=858.412}.  

\section{Conclusions}

{\tt SPYTHIA} simulates the physics of the MSSM using the
{\tt PYTHIA/JETSET} platform.  References to the underlying
physics that is simulated have been provided, as well as an
overview of the user interface to the program.  With the
appropriate choice of parameters, a wide range of MSSM
models can be studied, including those with a light
gravitino and extra $D$--term contributions to scalar masses.


\begin{thebibliography}{99}
\bibitem{susygen} {\tt SUSYGEN}~documentation~is~available~at~\hfill \\
vxcern::disk\$delphi:[katsanevas.susygen]susygen\_manual.ps.
\end{thebibliography}
\end{document}